\newcommand{\bk} {\mathbf k} 

\newcommand{\lp} {\lambda' }
\newcommand{\bkp}{ \mathbf k'} 
\newcommand{\epsb}[2]{{\boldsymbol{\epsilon}}( #1 ,\mathbf{#2})}
\newcommand{\epsu}[3]{\epsilon^ #1 ( #2, \mathbf{#3})}
\newcommand{\epsd}[3]{\epsilon_ #1 ( #2, \mathbf{#3})}
\newcommand{\chiu}[3]{\chi^ #1 ( #2, \mathbf{#3})}

\newcommand{\Epsu}[2]{\epsilon^ #1 (#2)}
\newcommand{\Epsd}[2]{\epsilon_ #1 (#2)}

\newcommand{\ad}{a^\dagger}
\newcommand{\ea}{\epsilon(\lambda)}
\newcommand{\eg}{\epsilon(\gamma)}

\newcommand{\adn}{a^{\dagger n}}
\newcommand{\adpm}[1]{a^{\dagger( #1 )}}
\newcommand{\vac}{|0\mspace{-5mu}>}
\newcommand{\pdu}[1]{ \partial^ { #1 }}
\newcommand{\pdd}[1]{ \partial_ { #1 }}
\newcommand{\DIV}[1]{ \partial^\nu{#1}_\nu}

\newcommand{\xii}{\lp}

\documentclass[10pt,a4paper]{article}
\usepackage[latin1]{inputenc}
\usepackage{amsmath}
\usepackage{amsfonts}
\usepackage{amssymb}
\begin{document}


\title{On the Covariant  Quantization of QED}

\author{Bernard R. Durney}
\date{2377 Route de Carc\'{e}s, F-83510 Lorgues, France durney@physics.arizona.edu} 

\maketitle

\smallskip

\smallskip\textbf{Abstract}
The commutation relations for bosons are field independent, and can be reliably 
inferred from the definition of creation and annihilation operators. Here, the
commutation relations are assumed known, and the quantum electrodynamics equations 
without sources are quantized with the unmodified Lagrangian. Non diagonal products of 
creation and annihilation operators of the form, cr(0)an(3)+ cr(3)an(0), where 0,3 denote respectively the time-like and longitudinal-polarizations, are present in both terms that
contribute to the Hamiltonian. However, the  contributions differ in sign, and therefore cancel.
In units of the photon's energy the coefficients of the Hamiltonian's four polarization states
 are -1/2, 1, 1, 1/2, clearly revealing the unphysical character of the time like and
 longitudinal polarization states. If the physical states are restricted to those that do not
contain unphysical polarization states, and if the Lorentz condition is satisfied, then the 
non diagonal terms of the field's momentum vanish, and both the Hamiltonian and Momentum are
 well behaved.\\
 A transformation of the basic vectors engenders in turn a transformation of the creation
  operators. The expression for the transformation that leaves invariant the commutation
 relations is derived.
 
 \smallskip \textbf{1. Commutation Relations}
Consider the action of an annihilation operator, $a ,$ on the state 
$\ad\ad\ad\vac,$ where $ \vac$ is the vacuum. The operator $a$ cannot just act on one
individual $\ad$ present in this state, it must instead act on each $\ad$
indiscriminately. Consequently we are led to write,  
$$
a(\ad\ad\ad) \sim \pm (a\ad)\ad\ad \pm \ad(a\ad)\ad \pm\ad \ad (a\ad),
\eqno{(1)}
$$
and hence in general,
$$
a(\ad)^{n}=\pm x_n \mspace{2mu} n \mspace{2mu}\adpm{n-1},
\eqno{(2)}
$$
where, the factor $a\ad$ has been included in $x_n$ which is 
unknown function of $n$. The following equalities are a consequence of Eq.(2) (we write
them for the plus sign only),
$$
\ad \{a\adn \} = x_n n \adn, { \  } a{\adpm{n+1}}= x_{n+1}(n+1)\adn
\eqno{(3a)}
$$
We assume that 
$ \ad \{ a \adn \} =\{\ad a\} \adn,$ and $a \adpm{n+1}= \{ a \ad \}\adn,$ divide by 
$x_n, x_{n+1},$ substract, and obtain,  
$$ 
a{\adpm{n+1}}/ x_{n+1}-\ad \{a\adn \} / x_n) = \adn
\eqno{(3b)}
$$
We write Eq.(3b) as follows,
$$
\left\{ a\ad / x_{n+1}-  \ad a / x_n \right\} \adn = \adn,
\eqno{(3c)}
$$
and conclude that the quantity in bracket must be equal to unity. This relation can only
ne meaningful if $x_{n+1} = x_{n},$ an arbitrary number, which by normalizing the creation annihilation operators, can be chosen to be $\pm 1.$ Therefore the commutation relations are,
$$
 a\ad -  \ad a = \pm 1
 \eqno{(4)}
$$
 \\
  Below we deal with annihilation and creation operators associated with the photon's
 polarization vectors, i.e., with products of the following form, $\ea a(\lambda),$ $\ea
 \ad(\lambda),$ where $\ea$ with $\lambda = 0,1,2,3$ are the time-like, the two transverse, 
 and the longitudinal polarization vectors chosen to form a complete orthonormal
set, i.e,
 $$
 \epsilon(\lambda) \cdot \epsilon(\gamma) = \epsilon ^{\mu}(\lambda)  
\epsilon_\mu (\gamma ) = \zeta_\lambda\delta_{\lambda \gamma} 
 \eqno{(5a)}
 $$
 $$
\sum_{\lambda}\zeta_\lambda  \epsilon ^{\mu}(\lambda) \epsilon ^{\nu}(\lambda)= g^{\mu\nu},
 \eqno{(5b)}  
 $$ 
where $\zeta_\lambda$ is the signature of the metric, namely $-1,1,1,1.$ If we apply Eq.(4)
to $\ea a(\lambda),\eg \ad(\gamma),$ we obtain,
$$
[\ea a(\lambda),\eg \ad(\gamma)] = \ea\cdot\eg \mspace{2mu}[a(\lambda),\ad(\gamma)] =
 \zeta_\lambda\delta_{\lambda \gamma}\mspace{2mu} [a(\lambda),\ad(\gamma)]=\pm 1
 \eqno{(6)}
$$
Because of Eq.(1) it appears natural to choose the positive sign in Eq.(6), and write the commutation relations as follows,
$$
[a(\lambda),\ad(\gamma)] = \zeta_\lambda\delta_{\lambda\gamma}
\eqno{(7)}
$$
Clearly, the derivation of Eq.(7) is not a model of rigor, nevertheless it is not an
 overstatement to argue that merely from the definition of creation and annihilation operators
for bosons it is possible to reliably infer their commutation relations (c.f. Weinberg, p.173).
\\
\smallskip
\textbf{2. Lagrangian}
\smallskip
 In the absence of a charge-current density, the QED's  Lagrangian density is given
 by, 
 $$ 
  \mathcal{L} = -\frac{1}{4} F^{\mu\nu}(x)F_{\mu\nu}(x) 
 = -\frac{1}{2}\pdu{\mu} A^\nu(x)\pdd{\mu}A_\nu(x) +\frac{1}{2}\pdu{\mu}
      A^\nu(x)\pdd{\nu}A_\mu(x)
 \eqno{(8) }
 $$
 where$ F^{\mu\nu}(x)= \pdu{\mu} A^\nu(x)-\pdu{\nu}A^\mu(x).$
  Because the conjugate momentum $\Pi^0 = \partial \mathcal{L}/ \partial\dot{A_0} $ vanishes,
   this Lagrangian is not suitable to derive Eqs.(6) from commutation relations between fields
    and conjugate momentum; to achieve this aim a modification of the Lagrangian becomes
  necessary.
It can be argued however that since these  relations are independently known, it is 
of interest to perform the following calculations: quantize the fields obeying
$$ 
\Box A^\mu (x)=0,
\eqno{(9a) }
$$ 
and use then these fields to calculate the energy-momentum four vector,
$$
c\mathcal{P^{\nu}} = \int \mathrm{d}^3\mathbf{x}N(\Pi^\mu\pdd{\nu}A_\mu - g^{\nu 0}\mathcal{L})
 = \int \mathrm{d}^3 \mathbf{x}N\{(\pdu{\mu}A^0 - \pdu{0}A^\mu)\pdd{\nu}A_\mu - g^{\nu0}
  \mathcal{L}\} 
 \eqno{(10) } 
$$ 
where $\mathcal{L}$  is the unmodified Lagrangian of Eq.(8), and  $N$ stands for 
normal ordering.  It could conceivably be argued (and rightly so) that the pertinent equation 
to quantify is not Eq.(9a) but rather the one displayed in Eq.(9b) below, which is the one
that can be derived from the Lagrangian, $\mathcal{L}$ . However a stringent criterion concerning the viability of these calculations exists: the Hamiltonian (and momentum components)
 contain non diagonal products of creation and annihilation operators, $(\ad(\lambda)a(\gamma),\lambda \# \gamma)$ where $\lambda, \gamma$ define the polarization and these terms must cancel.
$$
\Box A^\mu (x) - \pdu{\mu}(\DIV A(x)) = 0.
\eqno{(9b) }  
$$
\smallskip
\textbf{3. Expansions of the Fields}
\smallskip
The fields $A^\mu(x) = A^{+ \mu}(x) + A^{- \mu}(x)$ are expanded as follows 
(c.f. Mandl and Shaw, p.76),
   $$
  A^{+\mu}(x) =\sum _{\bk\lambda}G(\bk)\epsu{\mu}{\lambda}{\bk }a(\lambda,\bk)e^{-ikx},
  \mspace{5mu}
  A^{-\mu}(x) =\sum _{\bk\lambda}G(\bk)\epsu{\mu}{\lambda}{\bk }a^\dagger(\lambda,\bk)e^{ikx}
  \eqno{(11a)}
  $$
    $$
  G(\bk) = (\hbar c^2/2V\omega_{\bk})^{1/2}, \mspace{15mu }k^0 = \omega_{\bk}/c=|\bk|
   \eqno{(11b)}
  $$
    $$
  \epsilon(\lambda, \bk) =(\delta _{0\lambda}, \epsb{\lambda}{\bk }),\mspace{10mu}
 \epsb{0}{\bk} =(0,0,0),\mspace{10mu} \epsb{3}{\bk } = \bk/|\bk|,
  \eqno{(11c)}
 $$ 
 $$
\bk\cdot\epsb{1}{\bk}= \bk\cdot\epsb{2}{\bk} = 0, \mspace{15mu}\epsb{i}{\bk} 
\cdot\epsb{j}{\bk}= \delta_{ij}, \mspace{5mu}i,j = 1,2
 \eqno{(11d)}
 $$
 Above, the contravariant components of vectors are listed;  $\epsilon(\lambda,\bk)$ with
  $\lambda = 0,1,2,3$ are the time like, the two transverse, and the longitudinal polarization
   state respectively. Bold {\it epsilons} denote three-vectors, periodic boundary
   conditions are impose on the fields: $A(0,y,z,t) = A(L,y,z,t),....,$ and $V=L^3.$ In Eqs.(11), the vectors    
   $\epsilon(\lambda, \bk)$ form an orthonormal set. Notice that the vectors
    $((b^2+1)^{1/2}, -b \bk/|\bk|), \epsilon(1, \bk), \epsilon(2, \bk) $
   and $(-b,(b^2+1)^{1/2}\bk/|\bk|)$ with $b$ arbitrary also form an orthonormal set.
   However the {\it epsilons} must form a {\it complete} orthonormal set, i.e., satisfy
   Eq.(5b), which is only the case if $b=0.$ The proof of Eq.(5b) is straightforward.
  Completeness requires that a function $f(\lambda,\lambda')$ can be found such that,
  $$ \sum _{\lambda,\lambda'} \epsilon^\mu(\lambda)\epsilon^\nu(\lambda')
  f(\lambda,\lambda') = g^{\mu\nu} 
   $$
   Therefore, 
   $$
   g^{\mu\nu}\epsilon_{\mu}(\alpha)\epsilon_{\nu}(\beta) = \zeta_\alpha
   \delta_{\alpha\beta} = 
    \sum _{\lambda,\lambda',\mu,\nu} \epsilon_{\nu} 
     (\beta)\epsilon^\nu(\lambda')\epsilon_\mu(\alpha)\epsilon^\mu(\lambda)
     f(\lambda,\lambda')= \zeta_\alpha\zeta_\beta f(\alpha,\beta),
  $$
   and in consequence, $f(\alpha,\beta) =  \zeta_\beta  \delta_{\alpha\beta}.$
   The normally ordered terms from $-\mathcal{L}$ and $N(\Pi^\mu \pdd{\nu}A_\mu)$  that
    contribute to the energy-momentum four vector $c\mathcal{P},$ are,
  $$
   L1 = \frac{1}{2}\pdu{\mu}A^{-\nu}\pdd{\mu}A^{+}_{\nu} + \frac{1}{2}
   \pdd{\mu}A^{-}_{\nu}\pdu{\mu}A^{+\nu }= -\pdu{\mu}A^{-\nu}\pdd{\mu}A^{+}_{\nu }
   \eqno{(12a)}
   $$
   $$
   L2 = -\frac{1}{2}\pdu{\mu}A^{-\nu} \pdd{\nu}A^{+}_\mu - \frac{1}{2}\pdd{\nu}A^{-}_{\mu}
    \pdu{\mu}A^{+\nu} = -\pdu{\mu}A^{-\nu} \pdd{\nu}A^{+}_\mu
  \eqno{(12b)}
  $$
  $$
  P1 = \pdu{\mu}A^{-0}\pdd{\nu}A^{+}_{\mu} + \pdd{\nu}A^{-}_{\mu}\pdu{\mu}A^{+0}
  \eqno{(12c)}
  $$
  $$
  P2 =  - \pdu{0}A^{-\mu}\pdd{\nu}A^{+}_{\mu} -\pdd{\nu}A^{-}_{\mu}\pdu{0}A^{+\mu}.
  \eqno{(12d)}
   $$
  Of the momentum terms only $P1,$ and contain terms as $\ad(\lambda)a(\lambda')$ with 
  $\lambda \# \lambda'$ that are  non diagonal products of creation and annihilation operators. 
  $L1,$ when integrated over $x,$ is proportional to $k^2 = 0,$
  and therefore vanish. The substitution of expansions (11a) into Eqs(12) leads then to,
  $$
  L2= - \sum_{{\bk}{\bkp}{\lambda}{\xii}} G(\bk)G(\bkp)k^\mu k'_\nu\epsu{\nu}{\lambda} 
   {k}\epsd{\mu}{\lambda'}{k'}\ad(\lambda,\bk)a(\lambda',\bkp)e^{i(k-k')x}
  \eqno{(13a)}
  $$
  $$
  P1 =\sum_{{\bk}{\bk'}{\lambda}{\xii}} G(\bk)G(\bk')
  \Big\{ k^\mu k'_\nu \epsu{0}{\lambda}{k}\epsd{\mu}{\lambda'}{k'}+
   k_\nu  k'^\mu \epsd{\mu}{\lambda}{k}\epsu{0}{\lambda'}{k'} \Big\} \times
$$
$$
\ad(\lambda,\bk) a(\lambda',\bk')e^{i(k-k')x}
  \eqno{(13b)}
  $$
  $$
   P2 = -\sum_{{\bk}{\bk'}{\lambda}{\xii}} G(\bk)G(\bk')
   \Big \{ 
 k^0 k'_\nu \epsu{\mu}{\lambda}{k}\epsd{\mu}{\lambda'}{k'}
  + k_\nu k'^0 \epsd{\mu}{\lambda}{k}\epsu{\mu}{\lambda'}{k'}\Big\}\times
$$
$$
\ad(\lambda,\bk) a(\lambda',\bk')e^{i(k-k')x}
\eqno{(13c)}
 $$
The terms $L2, P1,$ and $P2$ contain the following products,
 $$k_\mu \epsu{\mu}{\lambda}{k}, \mspace{15mu } \epsd{\mu}{\lambda}{\bk}\epsu{\mu}{\lambda}{k}
 \eqno{(14)} 
 $$
 that are tabulated below,
 \begin{equation}{\nonumber}
\begin{matrix}
\lambda      & 0 & 1 & 2 &3  \\
\mspace{54mu} k_\mu \epsu{\mu}{\lambda}{k} &  k^0\zeta_0 & 0 & 0 & k^0\zeta_3 \\
\mspace{86mu} \epsilon_{\mu}(\lambda,\bk)\epsu{\mu}{\lambda}{k} & \zeta_0  & \zeta_1 & \zeta_2
 & \zeta_3   
\end{matrix}
\end{equation}
$\mspace{54mu}$ {\bf Table 1} Values of the above scalar products as a function of $\lambda.$
\\
\\
\smallskip
\textbf{4. The Hamiltonian}
\smallskip
It is important to evaluate the non diagonal terms of the type
$\ad(\lambda,\bk)a(\lambda',\bk),$ with $\lambda \# \lambda'$ present in Eqs. (13).
Because the integration over $\mathbf{x}$ introduces a delta function, ${\bk}' = \bk,$
 and $\bk$ will be omitted in the argument of creation-annihilation operators, as well as the factor $e^{i(k-k')x}.$ Consider the L2 term. It follows from Table 1 that,
$$
-\sum_{{\lambda}{\lambda'}}k_\nu\Epsu{\nu}{\lambda}\ad(\lambda)\ k^\mu\Epsd{\mu}{\lambda'}
 a (\lambda')= 
 $$
 $$-k_0^2 \sum_{{\lambda}{\lambda'}}
(\zeta_0 \delta_{0\lambda} +\zeta_3 \delta_{3\lambda}) (\zeta_0 \delta_{0\lambda'} +
  \zeta_3\delta_{3\lambda'}) \ad(\lambda) a(\lambda')
  \eqno{(15)}
$$
The non diagonal terms are,
$$  
  -k_0^2 \sum_{{\lambda}{\lambda'}}\Big \{ \ad(\lambda)\zeta_0 \delta_{0\lambda}\mspace{5mu}
  a(\lambda')\zeta_3\delta_{3\lambda'} +
\ad(\lambda)\zeta_3 \delta_{3\lambda}\mspace{5mu}
a(\lambda')\zeta_0 \delta_{0\lambda'} \Big\} =
$$
$$
 k_0^2 (\ad(0) a(3) + \ad(3)a(0))
 \eqno{(16)}
$$
Concerning P1, the non diagonal terms are (recall that, $\epsilon^0(\lambda) = \delta_{0\lambda}),$
$$
k^0 k_\nu(\ad(0)a(3) +\ad(3) a(0))
 \eqno{(17)}
$$
For the Hamiltonian $(k_\nu = -k^0,)$ the  non diagonal terms vanish. Clearly the reason
for this cancellation  cannot be accidental; instead it must indicate that 
some underlying physics is correctly described. Because a term from the Lagrangian
contributes to this cancellation, the momentum operator of the field contains
non diagonal terms.  We proceed now to calculate the diagonal terms contributing to the
 Hamiltonian, designated hereafter by $L2',P1', P2'.$  
For $k=k', \lambda= \lambda', $ it is found from Eqs.(15),(13b), and(13c) that,
$$
 L2' = - \sum_{{\bk}{\lambda}} k_0^2 G(\bk)^2(\delta_{0\lambda}
+\delta_{3\lambda})\ad(\lambda,\bk)a(\lambda,\bk)e^{i(k-k')x}
\eqno{(18a)}
$$
$$
 P1' = 2 \sum_{\bk} G(\bk)^2 k_0 k_\nu \ad(0,\bk)a(0,\bk)e^{i(k-k')x}
\eqno{(18b)}
$$
$$
 P2'  = 2 \sum_{{\bk}{\lambda}} G(\bk)^2 k_0 k_\nu \zeta_\lambda\ad(\lambda,\bk)
  a(\lambda,\bk)e^{i(k-k')x}
\eqno{(18c)}
$$
where $\bk$ has been reintroduced. The calculation of the Hamiltopnian and Momentum are now straightforward, it is found,
$$
\mathcal{H} = \int \mathrm{d}^3 \mathbf{x}(L2'+P1'+P2') =  \sum_{\bk}\hbar \omega_{\bk}
\Big \{-\frac{1}{2}\ad(0,\bk) a(0,\bk)+ \ad(1,\bk) a(1,\bk)
$$
$$
+\ad(2,\bk) a(2,\bk)+ \frac{1}{2}\ad(3,\bk) a(3,\bk) \Big \}, \mspace{10mu}
\hbar \omega_{\bk} = 2 k_0^2 L^3 G(\bk)^2
\eqno{(19a)}
$$
$$
\mathcal{P}^j = \sum_{\bk}\hbar k^j\Big \{-\frac{1}{2}\big(\ad(0,\bk) a(3,\bk)+
 \ad(3,\bk) a(0,\bk)\big ) +\ad(1,\bk) a(1,\bk)+
$$
$$ +\ad(2,\bk) a(2,\bk) +\ad(3,\bk) a(3,\bk) \Big \}, \mspace{10mu}
G(\bk) = (\hbar c^2/2V\omega_{\bk})^{1/2}
\eqno{(19b)}
$$
Notice that $\mathcal{P}^j$ does not contain $\ad(0,\bk) a(0,\bk),$ the contributions due
to $P1'$ and $P2'$ cancel. The integration over $\mathbf{x}$ of $L2'+P1'+P2'$ gives rise to 
the factor $(2\pi)^3 \delta^3(\bk-\bk') = L^3 \delta_{\bk\bk'}$ (c.f. Maggiore, p.84), and the
equalities $L^3 = V,$ and $k^0 =\omega_{\bk}/c$ lead then immediately to Eqs.(19).\\
 It is instructive to compare Eq.(19a) with Mandl and Shaw's Hamiltonian, namely,
 $\mathcal{H'},$ obtained with a modified Lagrangian (notice that the authors define
  $\zeta_r$ as {\it minus} the signature of the metric, and that their Eq.(5.32), p.79
  for $\mathcal{H'},$ unlike Eq.(20) below, is not metric independent, the factor $\zeta_3$
  being absent).
   $$
  \mathcal{H'}=  \zeta_3\sum_{{\bk} {\lambda}} \zeta_\lambda\hbar\omega_{\bk}
    \ad(\lambda,\bk)a(\lambda,\bk).
  \eqno{(20)}
  $$
  In Eq.(20), the longitudinal photon masquerades as a real photon, {\it not in Eq.(19a)}
  Furthermore the positive feature in Eq.(20), namely the opposite signs of the time like
  and longitudinal products of operators (c.f. Mandl and Shaw, p.80) {\it has been preserved 
  in Eq.(19a)}.{\it The Hamiltonian is indeed flawless}. Notice that for the physically
  meaningful polarizations the Hamiltonian and Momentum are compatible. The momentum operator, 
  however, contains non linear terms, issue that will be further discussed later on.  
  \\
\\
\smallskip
\textbf{5. Transformation properties of the annihilation and creation operators.}
\smallskip
$ \mspace{1mu}$ Assume that A Lorentz transformation is performed on the basic vectors,
$$
\chiu{\mu}{\lambda}{k} =\Lambda^\mu_\nu\epsu{\nu}{\lambda}{k}.
\eqno{(21)}
$$
In what follows "creation operators" is also meant to cover
annihilation operators. A transformation of the basic vectors, in turn engenders a
 transformation on the creation operators, functions of the known quantities, $\chiu{\mu}
{\lambda}{k}, \epsu{\mu}{\lambda}{k}$ and $\ad(\lambda,\bk).$ Our aim is to derive the
 expression for the transformed creation operators that leaves invariant the commutation
 relations, namely,
$$
[a(\lambda,\bk),\ad(\lambda',\bk')] = \zeta_\lambda\delta_{\lambda\lambda'}\delta_{\bk\bk'}
\eqno{(22)}
$$
It will be shown that if,
$$
S(\lambda,\alpha,\bk) = \chiu{\mu}{\lambda}{k}\epsd{\mu}{\alpha}{k}
\eqno{(23)}
$$
then, the $a'^{\dagger}(\alpha,\bk)$ defined by,
$$
a'^{\dagger}(\alpha,\bk) =
 \zeta_\alpha\sum_{\lambda}\ad(\lambda,\bk)S(\lambda,\alpha,\bk)
\eqno{(24)}
$$
do indeed satisfy Eq.(22). Eq.(24) shows that the commutation relations 
for the transformed operators can be written,
$$
[a'(\alpha,\bk),a'^{\dagger}(\beta,\bk')]=\zeta_\alpha\zeta_\beta\sum_{{\lambda}{\lambda'}}
[a(\lambda,\bk),a^{\dagger}(\lambda',\bk')]S(\lambda,\alpha,\bk) S(\lambda',\beta,\bk') 
$$
$$
=\zeta_\alpha\zeta_\beta\sum_{\lambda}\zeta_\lambda\epsu{\nu}{\lambda}{k}\epsu{\rho}{\lambda}{k}
\Lambda^{\mu}_{\nu}\Lambda^{\eta}_{\rho}
\epsd{\eta}{\beta}{k}\epsu{\rho}{\lambda}{k},
\eqno(25)
$$
where Eqs.(22) and (23) have been used. Because the basic vectors form a complete orthonormal
 set, we can replace the sum over  $\lambda$ by $g^{\nu\rho},$ which leads to,
$$
[a'(\alpha,\bk),a'^{\dagger}(\beta,\bk)]=\zeta_\alpha\zeta_\beta\Lambda^{\mu}_{\nu}
\Lambda^{\eta}_{\rho}g^{\nu\rho} \epsd{\mu}{\alpha}{k}\epsd{\eta}{\beta}{k}
= \zeta_\alpha\zeta_\beta g^{\mu\eta} \epsd{\mu}{\alpha}{k}\epsd{\eta}{\beta}{k}, 
\eqno{(26)}
$$
because for a Lorentz transformation, 
$$
\Lambda^{\mu}_{\nu} \Lambda^{\eta}_{\rho}g^{\nu\rho}= g^{\mu\eta}. 
\eqno{(27)}
$$
In consequence,
$$
[a'(\alpha,\bk),a'^{\dagger}(\beta,\bk)]=\zeta^2_\alpha\zeta_\beta\delta_{\alpha\beta},
\eqno{[28]}
$$
i.e., the commutation relations are indeed unchanged. This transformation of the creation
operators leaves invariant the $A^{\pm\mu}(x$) fields in Eq.(11a).\\\hspace*{20pt}An 
operator-equation as $a(3,\bk) - a(0,\bk) =0,$ needed to fulfil the Lorentz condition is not
 consistent with the commutation relations and is imposed instead, as suggested by Gupta and
 Bleuler,  on its product with the wave function, c.f. Maggiore, p.103. Notice that
 the transformed Lorentz condition is given by, $a'(3,\bk) - a'(0,\bk)= \sum_{\lambda}
 \Lambda^{\mu}_{\nu}\epsu{\nu}{\lambda}{k} a(\lambda,\bk)(\epsd{\mu}{3}{k} -
 \epsd{\mu}{0}{k})$  which is not a function only of $a(3,\bk) - a(0,\bk).$\\
Here, if we require that physical states satisfy the relation,
$$
\big( \ad (0,\bk)a(0,\bk)+ \ad(3,\bk)a(3,\bk)\big)|\varPsi \mspace{-9mu}>\mspace{5mu} = 0
\eqno{[29]}
$$ 
and add the Lorentz condition $\big(a(3,\bk) - a(0,\bk)\big)|\varPsi \mspace{-9mu}>\mspace{5mu}
 =0,$ then it is straightforward to show that the non diagonal terms in the field's momentum
vanish, i.e., 
$$
\big(\ad(0,\bk)a(3,\bk) +\ad(3,\bk) a(0,\bk)\big)|\varPsi \mspace{-9mu}>\mspace{5mu} = 0.
\eqno{[30]}
$$
If, it is imposed on the physical states the stronger condition,
$$
\ad (0,\bk)a(0,\bk)|\varPsi \mspace{-9mu}>\mspace{5mu} = \ad(3,\bk)a(3,\bk)|\varPsi \mspace{-9mu}>\mspace{5mu} = 0
$$
then both the Hamiltonian and Momentum are compatible.\\
\smallskip \textbf{6. Acknowledgements}\\ 
 I am deeply grateful to Dr. Gian Giudice for helpful comments.
\\ 
\smallskip \textbf{7. References}\\
\smallskip
Maggiore, M.: {\it A Modern Introduction to Quantum Field Theory}, Oxford
University Press, New York, (2005)\\
\smallskip
Mandl, F. and Shaw, G.: {\it Quantum Field Theory}, Wiley, United Kingdom, (2010)\\
\smallskip
Weinberg, S.: {\it The Quantum Theory of Fields} Vol 1, Cambridge University Press,
New York, (2005)
\end{document}